\documentclass[12pt]{article}
\usepackage{graphicx,amssymb,amsmath,amsthm,cite,cases}
\def\ds{\displaystyle}

\def\expin{e^{\imath \frac{2\pi(i-1)(n-1)}{M}}}

\def\qs{{q^\star}}
\def\kqs{k_{q^\star}}
\def\oj{\hat{\operatorname{J}}}

\def\be{\begin{equation}}
\def\ee{\end{equation}}
\def\ben{\begin{eqnarray}}
\def\een{\end{eqnarray}}

\def\D{\mathcal{D}}
\def\R{\mathbb{R}}
\def\C{\mathbb{C}}

\def\til{\tilde}

\def\vd{\mathbf{d}}

\def\vf{\mathbf{f}}
\def\vp{\mathbf{p}}

\def\vfk{\mathbf{f}^{\kq}}

\def\vB{\mathbf{b}}

\def\vW{\mathbf{w}}
\def\vWt{\til{\mathbf{w}}}

\def\vr{\mathbf{r}}
\def\vd{\mathbf{d}}

\def\kq{k_q}

\def\Nq{N_b}

\newcommand{\la}{\langle}
\newcommand{\ra}{\rangle}

\title{Sparse Representation of Gravitational Sound}
\author{Laura Rebollo-Neira\\
Mathematics Department\\
Aston University\\
B3 7ET, Birmingham, UK\\
\vspace{0.1cm}\\
A. Plastino\\
IFLP-CCT-Conicet\\
National University of La Plata\\
CC 727, 1900 La Plata, Argentina}
\begin{document}
\maketitle
\baselineskip = 1.5\baselineskip
\begin{abstract}
Gravitational Sound clips produced by the 
 Laser Interferometer Gravitational-Wave Observatory (LIGO) 
 and the Massachusetts Institute of Technology (MIT)  
 are considered within the particular context of data
 reduction. We advance a detailed procedure to this effect and 
 show that these types of signals can be 
approximated with high quality using  
{\it significantly fewer elementary components} than those required 
within the standard orthogonal basis framework. 
Furthermore, a measure a local sparsity is shown 
 to render meaningful information about  the  
 variation of a signal along time, 
by generating  a set of local 
sparsity values which is  
much  smaller than the dimension of the signal. 
This point is further illustrated by recourse to a 
more complex signal, generated by  Milde Science Communication to divulge Gravitational Sound in the form a  
ring tone. 
\end{abstract}
\section{Introduction}
In 1905 
 Henri Poincar\'{e} first suggested that accelerated
 masses in a relativistic field should produce gravitational waves  \cite{CGS16}. 
 The idea was magisterially pursued  by
 Einstein via his celebrated theory of general relativity. In 1918 he published his famous
quadrupole formula resulting from 
the calculation of the effect of gravitational waves
\cite{Eis18}. 
A century later,
the LIGO Scientific Collaboration and Virgo Collaboration 
published a paper about the gravitational radiation they had
detected on September 2015 \cite{LV15}.
Ever since scientists believe to
have entered in a new era of astronomy, whereby  the
universe will be studied by `its sound' 
\cite{LV16a,LV16b,BHB16,YYP16,OMK16}.  Gravitational Sound (GS) signals will then be here scrutinized with advanced techniques.

In the signal processing field, the problem of
finding a sparse approximation for a signal consists in
expressing the signal as a superposition of 
as few elementary  
components as possible, without  significantly affecting
the quality of the reconstruction. In 
 signal processing  applications 
 the approximation is carried out on a signal partition, 
i.e.,
by dividing the signal into small pieces and
 constructing the approximation
for  each of those pieces of data. Traditional
 techniques would carry out
the task using an orthogonal basis. However,
enormous improvements in sparsity can be achieved
using an adequate over-complete `dictionary' and
an appropriate mathematics method.
For the most part, these methods are based on minimization
of the $l_1$-norm \cite{CDS01} or are
greedy pursuit strategies
 \cite{MZ93, PRK93, Nat95, RNL02, ARNS04,
 Tro04, DTD06, NT09}, the latter being much more effective in practice.

Sparse signal representation of sound signals is a 
valuable tool for a number of auditory tasks
\cite{SL06,NHS12}. 
Moreover, the emerging theory of compressive sensing
\cite{Don06,CW08,Bar11} has
 enhanced the concept of sparsity by asserting that 
 the number of
measurements needed for accurate representation of
a signal informational content
decreases if the sparsity of the representation improves.
Hence, when some GS tones made 
with the observed Gravitation Wave (GW)
   were released, 
we felt motivated to produce a sparse approximation of
those clips. 

We simply analyze  the GS tones  from a processing 
viewpoint, regardless on how and why they have been 
generated. We consider a) a short tone made with the 
chirp {\tt{gw151226}} that has been detected, 
b) the theoretical  
  simulated theoretical GS,
 {\tt{iota\_20\_10000\_4\_4\_90\_h}}, and 
c) the {\tt{Black\_Hole\_Billiards}} ring tone, which is a 
more complex signal produced by superposition with an 
ad hoc independent percussive sound. 
The ensuing results are certainly interesting. If, in the future,  
 GS  signals are to be generated at large scale (as astronomical images 
have been produced \cite{hubweb,esoweb}), it is 
important to have tools for all kinds of processing of 
 those signals. 

{\it {The central goal of 
this Communication is to present evidences of the 
significant gain in sparsity achieved if a GS signal 
is approximated with high quality outside the orthogonal basis framework}}. 
For demonstration purposes we have made available 
the MATLAB routines for implementation of the method. 
\label{Intro}
\section{Some Preliminary Considerations}
The traditional frequency decomposition of a signal  
given by $N$ sample points, $f(i),\,i=1,\ldots,N$,
involves the Fourier expansion 
$ f(i) =\frac{1}{\sqrt{N}} \sum_{n=1}^M c(n) \expin, \quad i=1,\ldots,N. $
The values $|c(n)|,\, n=1,\ldots,M=N$ are called
the discrete Fourier spectrum of the signal, and can be
evaluated in a very effective manner via the
Fast Fourier Transform (FFT). 
For $M>N$ even if the coefficients in the above 
expansion can still be calculated via FFT, by zero padding, 
these are not longer unique. Finding 
a sparse solution is the goal of 
 sparse approximation techniques.  

The problem of the sparse approximation of 
a signal, outside the orthogonal basis setting, 
consists in using elements of a redundant set, 
 called a  {\it  dictionary}, 
 for constructing 
an  approximation involving  a number of elementary 
components which is significantly smaller than the signal 
dimension. For signals whose structure varies with time, 
 sparsity performs better when the 
approximation is carried out on a signal partition.
In order to give precise definitions we  
 introduce at this point the notational usual conventions: 
$\R$ and $\C$
represent the sets of real and complex
and numbers, respectively.
Boldface fonts are used to indicate Euclidean vectors
 and standard mathematical fonts  to
indicate components,  e.g., $\vd \in \C^N$ is a vector of
$N$-components
$d(i) \in \C^N\,, i=1,\ldots,N$.
The operation
$\la \cdot,\cdot \ra$ indicates the Euclidean inner
product and $\| \cdot \|$ the induced norm, i.e.
$\| \vd \|^2= \la \vd, \vd \ra$, with the usual
inner product definition: For $\vd \in \C^N$
and $\vf \in \C^N$
$
\la \vf, \vd \ra = \sum_{i=1}^N f(i) d^\ast\!(i),
$
where $d^\ast\!(i)$ stands for the complex conjugate of
$d(i)$.

A partition of a signal $\vf \in \R^N$ 
is represented as a set of disjoint pieces,
 $\vf_q \in \R^{\Nq},\, 
q=1,\ldots,Q$, henceforth to be called `blocks',  
 which,  without loss of generality, are assumed  to
be all of the same size and such that $Q \Nq =N$.
 Denoting by
$\oj$ the concatenation operator, the
signal $\vf \in \R^N$ is `assembled' from the blocks as
$\vf=\oj_{q=1}^Q \vf_q$. This operation implies that
the first $N_1$ components of the vector $\vf$ are given
by the vector $\vf_1$, the next $N_2$ components by the
vector $\vf_2$, and so on.

A {\em{dictionary}} for $\R^{\Nq}$ is
 an {\em{over-complete}} set of (normalized to unity)
elements
$\D=\{\vd_n \in \R^{\Nq}\,; \| \vd_n\|=1\}_{n=1}^M,$
 which are called {\em{atoms}}.

\section{Sparse Signal Approximation}
Given a signal partition $\vf_q \in \R^{\Nq},\, q=1,\ldots,Q$
and a dictionary $\D$, the $\kq$-term approximation
for each block is given by an atomic decomposition 
of the form
\be
\label{atoq}
\vfk_q= \sum_{n=1}^{\kq}
c^{k_q}(n) \vd_{\ell^{q}_n},
\quad q=1,\ldots, Q. 
\ee
The approximation to the whole signal is then  
obtained simply by joining the approximation for 
the blocks as
$\vf^K= \oj_{q=1}^Q \vfk_q,$
where $K= \sum_{q=1}^Q \kq$.

\subsection{The Method}
The problem of finding the minimum number 
of $K$ terms 
such that $\|\vf - \vf^K\| <\rho$, for a 
given tolerance parameter $\rho$, 
is an NP-hard problem \cite{Nat95}.
In practical applications, one looks 
for tractable sparse solutions. For this purpose 
we consider the  
Optimized Hierarchical Block Wise (HBW) version \cite{LRN16} of the
Optimized Orthogonal Matching Pursuit (OOMP) \cite{RNL02}
 approach. This entails that, in addition to 
selecting the dictionary atoms for the approximation 
of each block, the blocks are ranked for their sequential 
stepwise approximation. As a consequence, the approach 
is optimized in the sense of 
minimizing, at each iteration step, the norm of 
the total residual error $\|\vf - \vf^K\|$ \cite{LRN16}. 
As will be illustrated in Sec.~\ref{NE}, when approximating 
 a signal with pronounced amplitude variations  
the sparsity result achieved by this 
 strategy is remarkable superior to that  
  arising when the approximation of 
each block is completed at once, i.e., when the 
 ranking of blocks is omitted.
The OHBW-OOMP method is implemented using  the 
steps indicated below. 

{\bf{OHBW-OOMP Algorithm}}
\begin{itemize}
\item[1)]For $q=1,\ldots,Q$ 
initialize the algorithm by setting:
$\vr_q^0=\vf_q$, $ \vf_q^0=0$, $\Gamma_q= \emptyset$  
$\kq=0$, and  
selecting the `potential' first atom for
 the atomic decomposition of every
block $q$ as the one corresponding to the indexes  
$\ell_{1}^q$ such that
\be
\ell_{1}^q=\operatorname*{arg\,max}_{n=1,\ldots,M}
 \left |\la \vd_n,\vr_{q}^{\kq}\ra \right|^2,  
 \quad q=1,\ldots,Q.
\ee
Assign $\vW_1^q=\vB_1^q=\vd_{\ell_{1}^q}$.
\item[2)]Use the
OHBW  criterion  for
selecting the block to 
upgrade the atomic decomposition 
by adding one atom
\be
\label{hbwoomp}
q^\star=
\operatorname*{arg\,max}_{q=1,\ldots,q}
 \frac{|\la \vW_{\kq+1}^q, \vf_{q}
\ra|^2}{\|\vW_{\kq+1}^q\|^2}.
\ee
If $k_{\qs}>0$ upgrade vectors 
$\{\vB_n^{\kqs,\qs}\}_{n=1}^{\kqs}$ for block $\qs$ as
\be
\begin{split}
\label{BW}
\vB_{n}^{{\kqs}+1,\qs}&= \vB_{n}^{{\kqs},\qs} - \vB_{\kqs+1}^{{\kqs}+1,\qs}\la \vd_{\ell_{{\kqs}+1}}^{\qs}, \vB_{n}^{\kqs+1,\qs}\ra,\quad n=1,\ldots,\kq,\\
\vB_{\kqs+1}^{\kqs+1,\qs}&= \frac{\vW_{\kqs+1}^\qs}{\| \vW_{\kqs+1}^\qs\|^2}.
\end{split}
\ee
\item[3)] 
Calculate
\ben
\vr_{\qs}^{\kqs+1} &=& \vr_{q}^{\kqs} - \la \vW_{\kqs+1}^{\qs}, \vf_{\qs} \ra  \frac{\vW_{\kqs+1}^{\qs}}{\| \vW_{\kqs+1}^{\qs}\|^2},\nonumber \\ 
\vf_{\qs}^{\kqs+1} &=& \vf_{\qs}^{\kqs+1} + 
\la \vW_{\kqs+1}^{\qs}, \vf_{\qs} \ra \frac{\vW_{\kqs+1}^{\qs}}{\| \vW_{\kqs+1}^{\qs}\|^2}.
\een
Upgrade the set $\Gamma_{\qs} \leftarrow  \Gamma_{\qs} \cup 
\ell_{\kqs+1}$ and increase $\kqs\leftarrow \kqs +1$. 
\item[4)]
Select a new potential atom for the
 atomic decomposition of block $\qs$, using 
the OOMP criterion, 
i.e., choose $\ell_{\kqs+1}^q$ such that
\be
\label{oomp}
\ell_{\kqs+1}^q=\operatorname*{arg\,max}_{\substack{n=1,\ldots,M\\ n\notin \Gamma_q}}
 \frac{|\la \vd_n,\vr_{\qs}^{\kq}
\ra|^2}{1 - \sum_{i=1}^{\kq}
|\la \vd_n ,\vWt_i^q\ra|^2}, 
, \quad  \text{with} \quad \vWt_i^{\qs}= \frac{\vWt_i^{\qs}}{\|\vWt_i^{\qs}\|},
\ee
\item[5)]
Compute the corresponding new vector $\vW_{\kqs+1}^{\qs}$ as
\be
\begin{split}
\label{GS}
\vW_{\kqs+1}^q= \vd_{\ell_{\kqs+1}}^q - \sum_{n=1}^{\kqs} \frac{\vW_n^\qs}
{\|\vW_n^\qs\|^2} \la \vW_n^\qs, \vd_{\ell_{\kqs}}^q\ra.
\end{split}
\ee
including, for numerical accuracy,  the 
re-orthogonalizing step:
\be
\label{RGS}
\vW_{\kqs+1}^q \leftarrow \vW_{\kqs+1}^q- \sum_{n=1}^{\kqs} \frac{\vW_{n}^\qs}{\|\vW_n^\qs\|^2}
\la \vW_{n}^\qs , \vW_{\kqs+1}^q\ra.
\ee
\item[6)]Check if, for a given
$K$ and $\rho$ either the condition $\sum_{q=1}^Q \kq=K+1$  
or $\| \vf - \vf^K\| < \rho$ has been met. If 
that is the case, for $q=1,\ldots,Q$ compute the coefficients 
$c^{k_q}(n) = \la \vB_n^{\kq}, \vf_q \ra,\, n=1,\ldots, \kq$. 
Otherwise repeat steps 2) - 5).
\end{itemize}
{\bf{Remark 1:}} For all the values of $q$, 
the OOMP criterion \eqref{oomp} in the 
algorithm above ensures that, fixing the set 
of previously 
selected atoms, the atom corresponding to the 
indexes given by \eqref{oomp} minimizes the local  
residual norm $\|\vf_q -\vf_q^{\kq}\|$ \cite{RNL02}.
 Moreover, the OHBW-OOMP criterion \eqref{hbwoomp}, 
for choosing the block to upgrade the approximation,
 ensures the 
minimization of the total residual norm \cite{LRN16}.
Let us recall that the 
OOMP approach optimizes the Orthogonal Matching Pursuit 
(OMP) one \cite{PRK93}. The latter is also 
an optimization of the plain Matching Pursuit (MP) 
method \cite{MZ93}(see the discussion in \cite{RNL02}).

\subsection{The Dictionary}
The degree of success in achieving high sparsity
using a dictionary approach depends on both, 
the suitability of the mathematical method for 
finding a tractable sparse solution and the dictionary 
itself.
As in the case of melodic music \cite{LRN16,RNA16}, we found 
the trigonometric dictionary $\mathcal{D}_T$, 
which is the union of the dictionaries $\mathcal{D}_{C}$ and 
$\mathcal{D}_{S}$ given below, to be an 
appropriate dictionary for approximating these
GS signals. 
\ben
\mathcal{D}_{C}^x&=&\{w_c(n)
\cos{\frac{{\pi(2i-1)(n-1)}}{2M}},i=1,\ldots,\Nq\}_{n=1}^{M}\nonumber\\
\mathcal{D}_{S}^x&=&\{w_s(n)\sin{\frac{{\pi(2i-1)(n)}}{2M}},i=1,\ldots,\Nq\}_{n=1}^{M}.\nonumber
\een
In the above sets $w_c(n)$ and $w_s(n),\, n=1,\ldots,M$
are normalization factors.

\begin{figure}[!ht]
\begin{center}
\includegraphics[width=9cm]{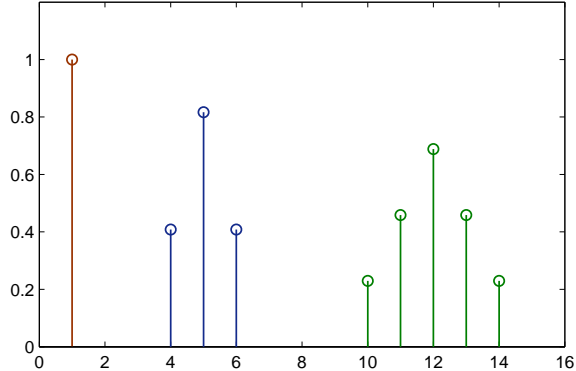}
\caption{\small{Prototype atoms $\vp_1, \vp_2$ and 
$\vp_3$, which generate the
dictionaries $\mathcal{D}_{P1}$, $\mathcal{D}_{P1}$  
and $\mathcal{D}_{P3}$ by sequential
translations of one point. Each prototype is shown in a
different color. \label{pato}}}
\end{center}
\end{figure}

We also found that 
 sparsity may benefit by the inclusion of 
a dictionary which is constructed
by translation of the prototype atoms, $\vp_1, \vp_2$ and 
$\vp_3$
in Fig.~\ref{pato}. Denoting  by 
$\mathcal{D}_{P_1}$, $\mathcal{D}_{P_2}$ 
and $\mathcal{D}_{P_3}$ the
dictionaries  arising  by translations of the atoms 
$\vp_1$, $\vp_2$,  and $\vp_3$, respectively, 
the dictionary $\mathcal{D}_{P}$ is 
built as
$\mathcal{D}_{P}= \mathcal{D}_{P_1} \cup \mathcal{D}_{P_2}
\cup \mathcal{D}_{P_3}$. 
The whole mixed dictionary is then  
$\mathcal{D}_M = \mathcal{D}_{T} \cup 
\mathcal{D}_{P}$, with 
$\mathcal{D}_{T}= \mathcal{D}_{C} \cup \mathcal{D}_{S}$. 
Interestingly enough, the dictionary  $\mathcal{D}_M$ 
happens to be a sub-dictionary of a larger dictionary 
proposed in \cite{RNB13} for producing sparse 
representations of astronomical images, the difference 
being that, in this case, sparsity does not improve 
in a significant way by further enlarging the dictionary.

From a computational  viewpoint  
the particularity of the sub-dictionaries $\mathcal{D}_{C}$
and $\mathcal{D}_{S}$ is that  
the inner product with 
all its elements can be evaluated via FFT. This 
  possibility reduces the complexity
of the numerical calculations when the partition 
unit $\Nq$ is large \cite{LRN16, RNA16}. 
%
Also, the inner products
 with the atoms of the  dictionaries 
 $\D_{P_2}$  and $\D_{P3}$
can be effectively   implemented,  
all at once,  via a convolution operation.\\
{\bf{Note:}} 
The MATLAB routine implementing the OHBW-OOMP approach, 
dedicated to the dictionary introduced 
in this section, has been 
made available on \cite{paperpage}.

\subsection{The Processing}
\label{NE}
We process now the three signals we are considering here:
\begin{itemize}
\item[a)]The audio representation of the 
detected {\tt{gw151226}} chirp \cite{LigoData}.
\item[b)]
The tone of the theoretical gravitational 
wave {\tt{iota\_20\_10000\_4\_4\_90\_h}} \cite{MITData}.
\item[c)]The {\tt{Black\_Hole\_Billiards}} ring tone 
\cite{LigoData}.
\end{itemize}
The quality of an approximation is measured by the
Signal to Noise Ratio (SNR) which is defined as
\be
\text{SNR}=10 \log_{10} \frac{\| \vf\|^2}{\|\vf - \vf^K\|^2}=
10 \log_{10}\frac{\sum_{\substack{i=1\\q=1}}^{\Nq,Q} |f_q(i)|^2}
{\sum_{\substack{i=1\\q=1}}^{\Nq,Q} |f_q(i) -f^{\kq}_q(i)|^2}.
\ee
The sparsity of the whole representation is measured by
 the Sparsity Ratio (SR) defined as
$\ds{\text{SR}= \frac{N}{K}}$, where $K$ is the total
number of coefficients in the signal representation as
defined above.

\subsubsection*{Audio representation of the chirp
 {\tt{gw151226}}}
\begin{figure}[!ht]
\begin{center}
\includegraphics[width=8cm]{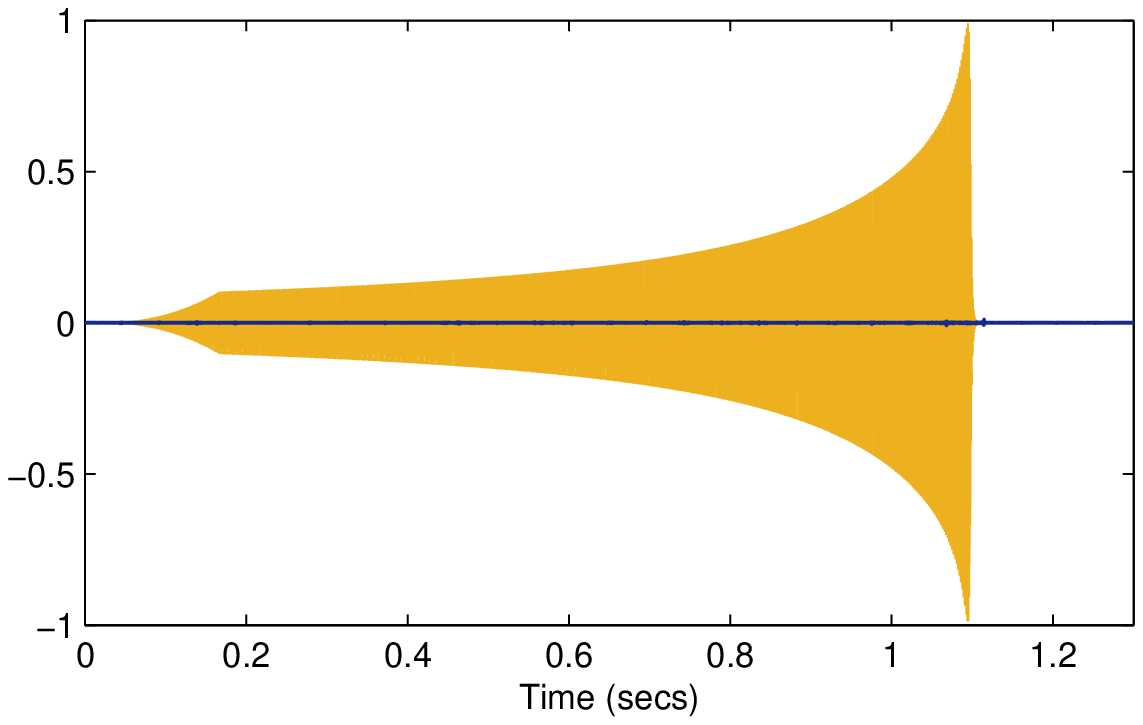}
\includegraphics[width=8cm]{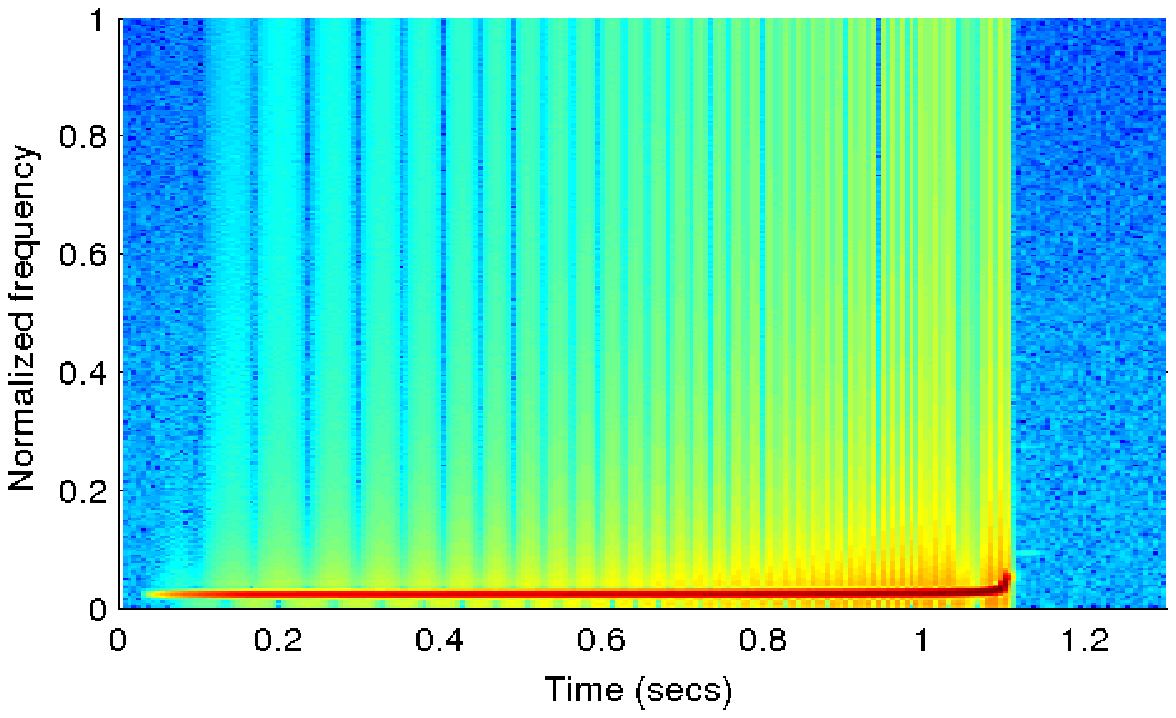}
\caption{\small{The graph on the left represents the 
 clip {\tt{gw151226}}.  
 The central line  
is the difference between the approximation, up to 
SNR=50dB, and the signal. The right graph is the 
classic spectrogram of the clip on the left. \label{gwc}}}
\end{center}
\end{figure}

This clip, made with the detected short 
chirp {\tt{gw151226}},   
is plotted in the left graph of
Fig.\ref{gwc}. The graph on the right is its 
classic spectrogram. 
When an orthogonal basis for approximating 
 these signals is used, the best 
sparsity result is achieved with the 
Discrete Cosine Transform (DCT). Hence, 
we first approximate this clip, up to SNR=50dB,
 by nonlinear thresholding of the DCT coefficients. 
The best SR (SR= 28.7) is obtained for $\Nq=N= 65536$, 
i.e., by processing the  signal as 
a single block. Contrarily, when approximating the clip 
 using the trigonometric dictionary $\D_{T}$,
 the best result is obtained for $\Nq=2048$, 
 achieving a much higher SR. Approximating 
each block at once, with the OOMP approach, SR=209.4, and 
raking the blocks with the  OHBW-OOMP approach SR= 263.2.
 Let us stress that this implies a gain in 
sparsity result of $817\%$  with respect to the DCT 
approach for the same value of SNR.
 The central dark line
in the left graph of Fig.~\ref{gwc} represents the
difference between the signal and its approximation, up to
SNR=50dB. For this chirp the inclusion of the 
dictionary $\D_{P}$ would not improve sparsity. 

\subsubsection*{Theoretical Gravitational Wave Sound}

This is the  {\tt{iota\_20\_10000\_4\_4\_90\_h}}
 gravitational wave, 
 which belongs to the family of Extreme Mass Ratio Inspirals
\cite{Hug00,Hug01,GHK02,HDF05, DH06} available on 
 \cite{MITData}.
\begin{figure}[!ht]
\begin{center}
\includegraphics[width=8cm]{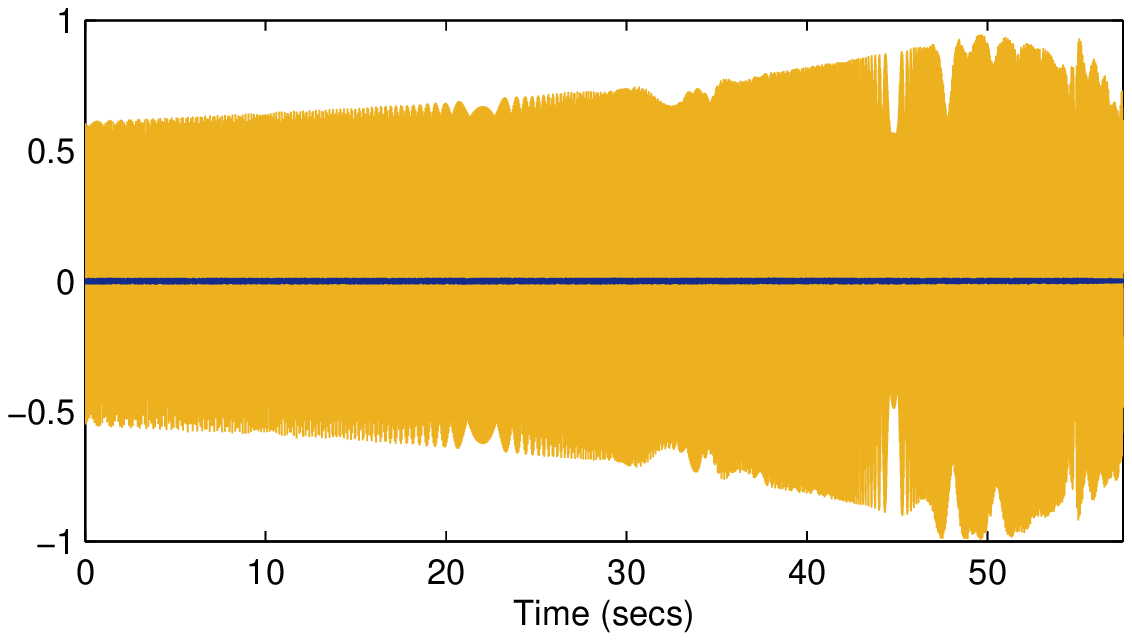}
\includegraphics[width=8cm]{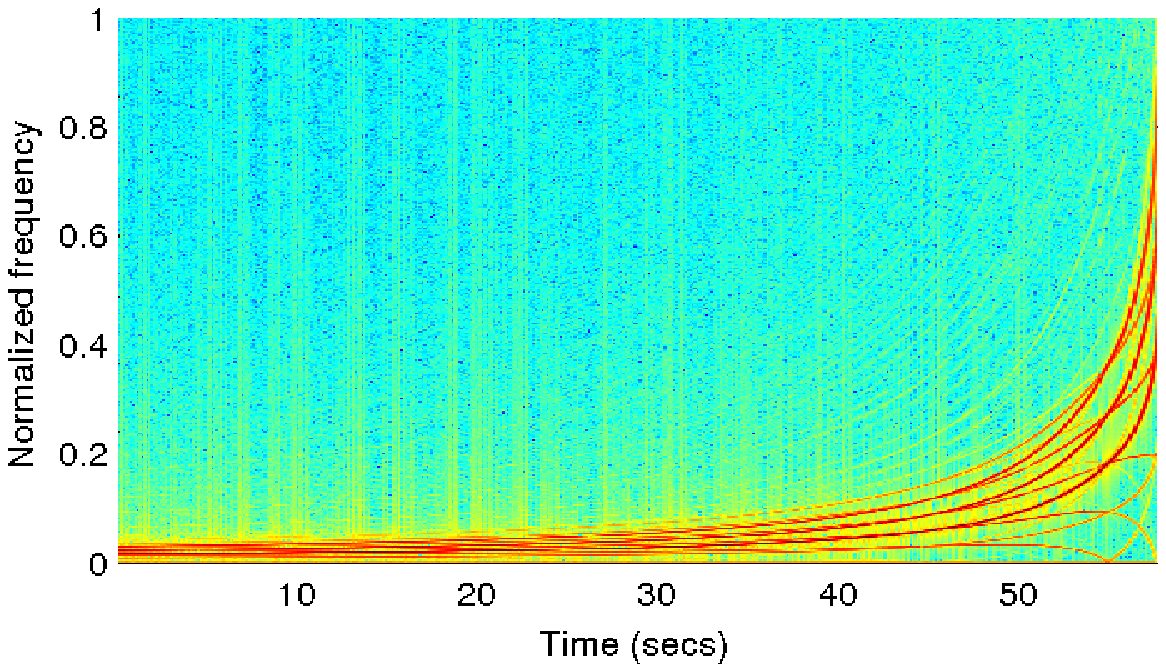}
\caption{\small{The graph on the left represents the
  {\tt{iota\_20\_10000\_4\_4\_90\_h}}
 tone. 
 The central line 
is the difference between the approximation, up to 
SNR=50dB, and the signal. The right graph is the 
spectrogram of the clip on the left. \label{iota}}}
\end{center}
\end{figure}
It consists of $N=458752$ data points plotted 
in the left graph of Fig.~\ref{iota}. The graph on the 
right is its classic spectrogram. In this case the  best 
SR result (SR=5.1),  produced by 
nonlinear thresholding of
the DCT coefficients for approximating the signal up to 
SNR=50dB, is obtained with $\Nq=16384$. A much smaller 
value of $\Nq$ ($\Nq=2048$) is required  to achieve   the 
   best SR result  (SR=10.8) 
with the OHBW-OOMP method and the trigonometric 
dictionary. With the mixed dictionary $\D_M$  there is 
a further improvement: SR=11.9. 
The central dark line in the left graph of Fig.~\ref{iota} represents the
difference between the signal and its approximation, up to
SNR=50dB. For this signal the gain in SR with respect to 
the DCT approximation is $136\%$. Since the amplitude of the 
signal  does not vary much along time, the SR 
obtained by approximating each block at once, with OOMP, does not 
 significantly differ  from the values obtained applying the 
OHBW-OOMP strategy. 

\subsubsection*{The {\tt{Black\_Hole\_Billiards}} ring tone}

\begin{figure}[!ht]
\begin{center}
\includegraphics[width=8cm]{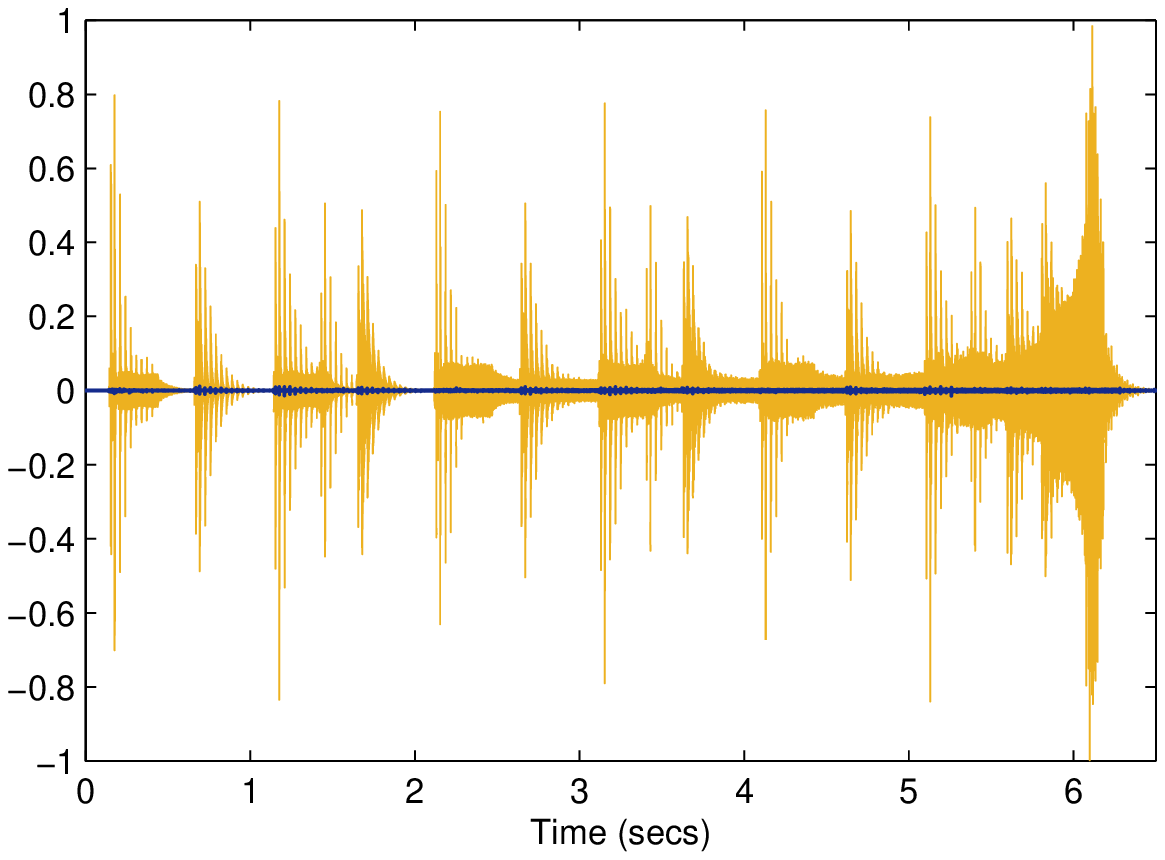}
\includegraphics[width=8cm]{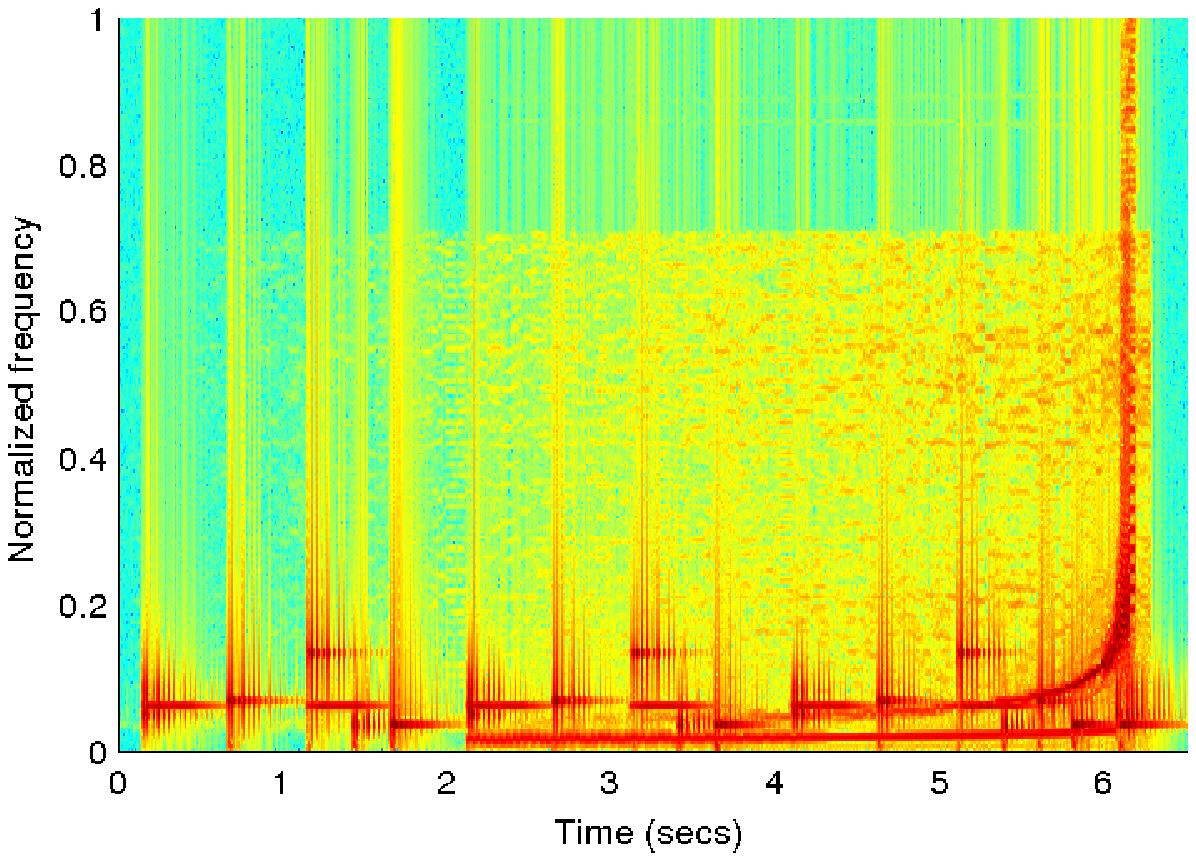}
\caption{\small{The graph on the left represents the
  {\tt{Black\_Hole\_Billiards}} clip. Credit:
Milde Science Communication.
 The central dark  line
is the difference between the approximation, up to
SNR=40dB, and the signal. The right graph is the
spectrogram of the clip on the left. \label{bhb}}}
\end{center}
\end{figure}

In order to stress the relevance of the technique 
 for representing features 
of more complex signals using a very reduced set of points, 
we consider here 
 {\tt{Black\_Hole\_Billiards}} ring tone  available on 
 \cite{LigoData}. This clip was created by 
Milde Science Communication by superimposing 
a sound of percussive nature (the billiards sound) to 
a GW chirp. It consisting of $N=262144$ samples 
 plotted in the left graph of Fig.~\ref{bhb}. 

The graph on the right is its classic spectrogram.
When processing the signal with DCT the 
best sparsity result when the approximation is carried out 
block by block up to the same error is SR=4.2, for  
SNR=40dB, and  corresponds to $\Nq=16384$. However, 
with $\Nq=2048$ the OHBW version for selecting DCT 
coefficients improves in this case the standard DCT result, 
  attaining SR=6.2. 
For an approximation of the same quality (SNR=40 dB) 
 the SR rendered by the OHBW-OOMP method 
with $\Nq=512$ and the
trigonometric dictionary $\D_T$ is SR= 12.1. 
With the mixed dictionary $\D_M$ this value increases to
SR=13.7.
The central dark line in the left graph of Fig.~\ref{bhb}
represents the
difference between the signal and its approximation, up to
SNR=40dB. It is worth commenting that, if with the same 
dictionary, the approximation were carried out without
ranking the blocks, i.e., approximating each block at 
once up to the same SNR, the value of SR would be 
only 6.7. This example highlights the importance of 
adopting the OHBW strategy for constructing the signal 
approximation, when the signal amplitude varies significantly along the domain of definition. 
 
\subsection{The Role of Local Sparsity}
The SR is a global measure of sparsity indicating the 
number of elementary 
 components contained in the whole signal. 
An interesting description of a the signal variation 
is rendered by a local measure of sparsity. For this 
we consider the local sparsity ratio
$sr(q)= \frac{\Nq}{\kq},\,q=1,\ldots,Q$ where, 
as defined above, 
$\kq$ is the number of coefficients in the
decomposition of the $q$-block and $\Nq$ the size of
the block.
\begin{figure}[!ht]
\begin{center}
\includegraphics[width=8cm]{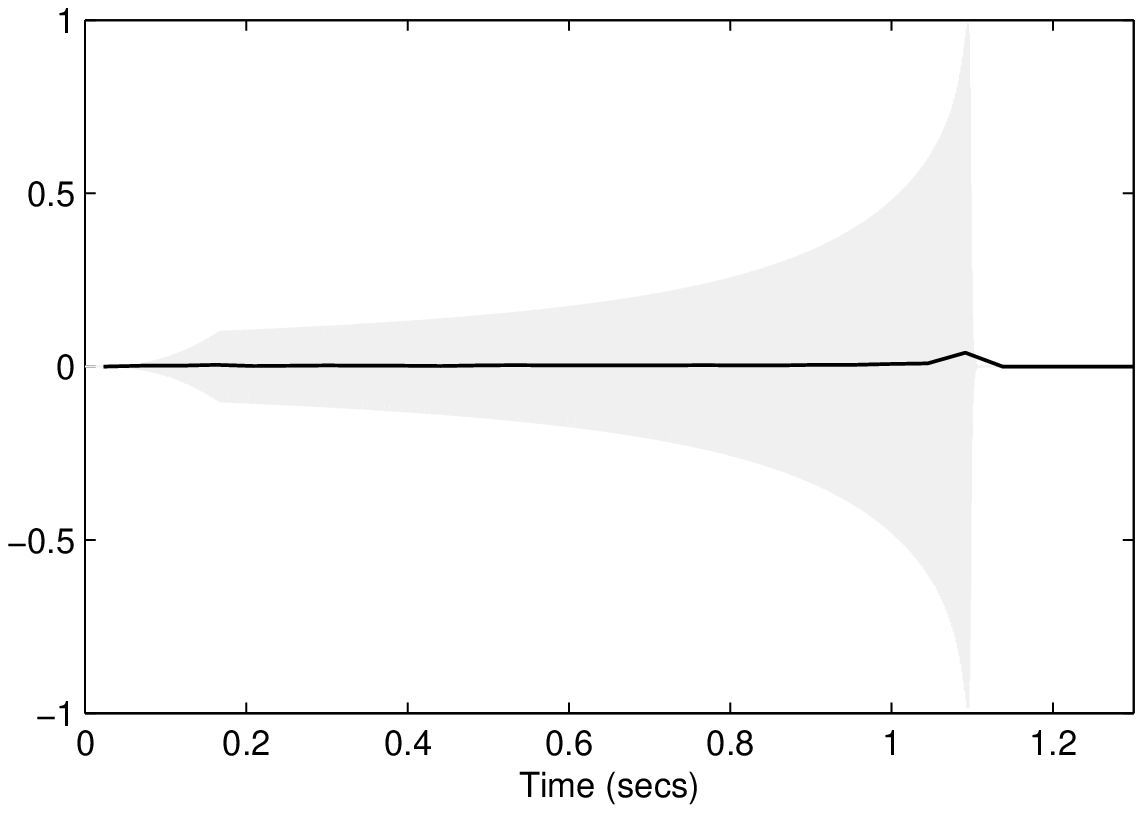}
\includegraphics[width=8cm]{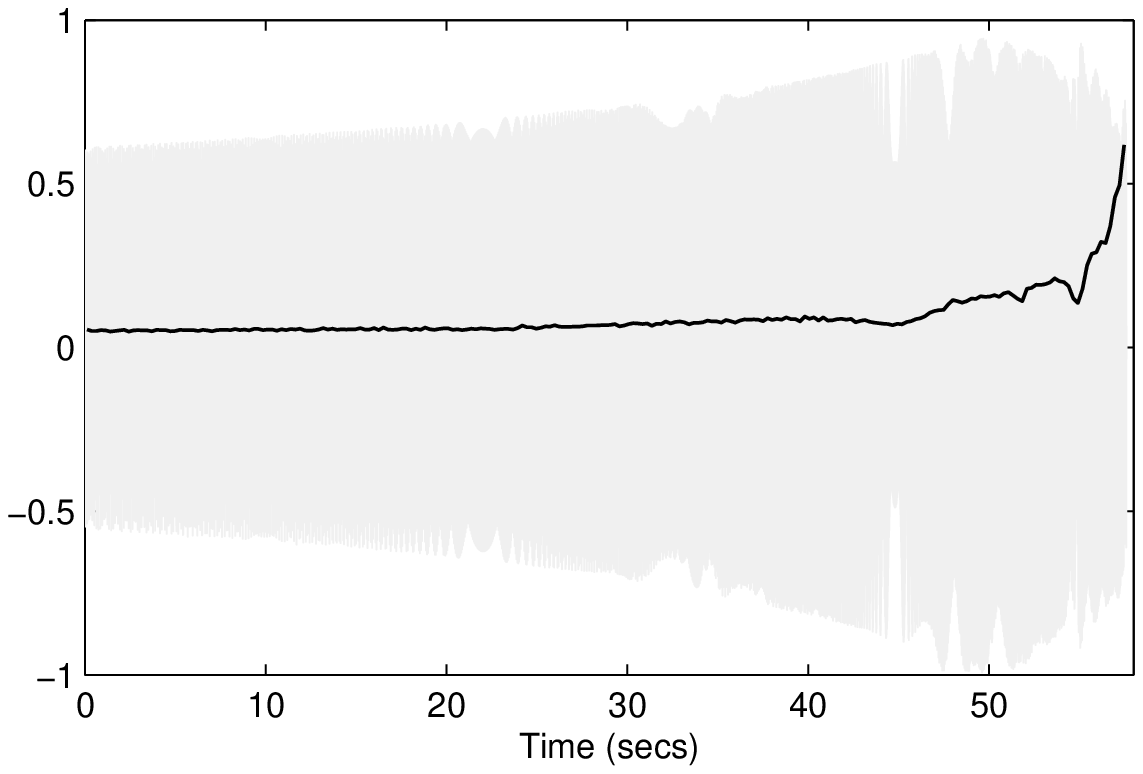}
\caption{\small{The dark line in the left graph  
joins the inverse local sparsity 
values for the clip {\tt{gw151226}}.
The right graph has the same description but for the 
{\tt{iota\_20\_10000\_4\_4\_90\_h}} clip.
 tone. \label{lsr}}} 
\end{center}
\end{figure}

 For illustration's convenience the dark 
line in both graphs of Fig.~\ref{lsr}
depicts the inverse of this local measure. This line 
 joins the values 
$1/sr(q),\, q=1,\ldots,Q$. 
Each of these values is located in the horizontal axis at
the center of the corresponding block and provides 
much information about the signal.  Certainly, 
simply from the observation of the the dark line 
in the left 
graph of Fig.~\ref{lsr}  (joining 32  points of 
inverse local sparsity ratio) one can realize 
 that the number of internal components in 
the clip  {\tt{gw151226}} is roughly 
constant along audiable part of the signal, 
 with a significant higher value only at the 
very end if this part.
 In the case
of the {\tt{iota\_20\_10000\_4\_4\_90\_h}}
 clip (right graph in the 
same figure) the line joining the 224 points of 
the inverse local sparsity ratio  indicates a  clear 
 drop of sparsity 
towards the end of the signal, where the rapid rise of the 
tone does occur (c.f. spectrogram in Fig.~\ref{iota}).

\begin{figure}[!ht]
\begin{center}
\includegraphics[width=8cm]{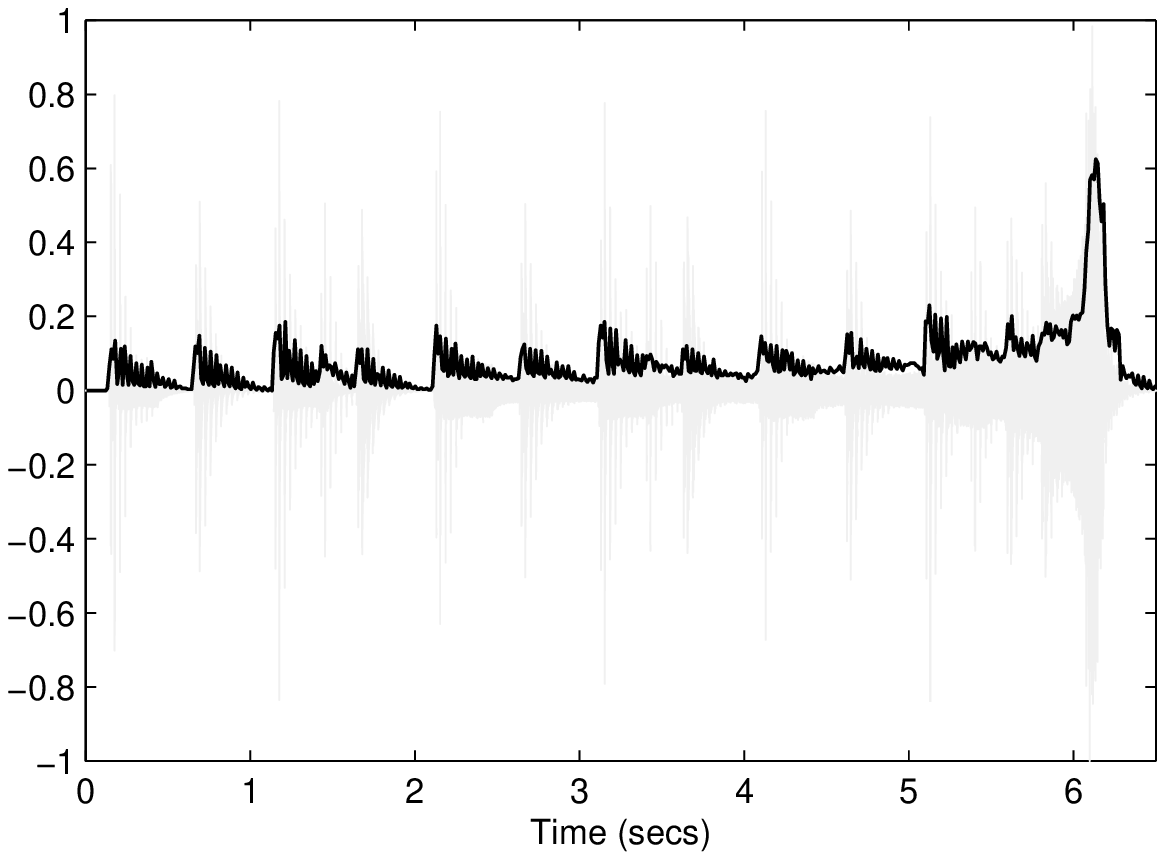}
\includegraphics[width=8cm]{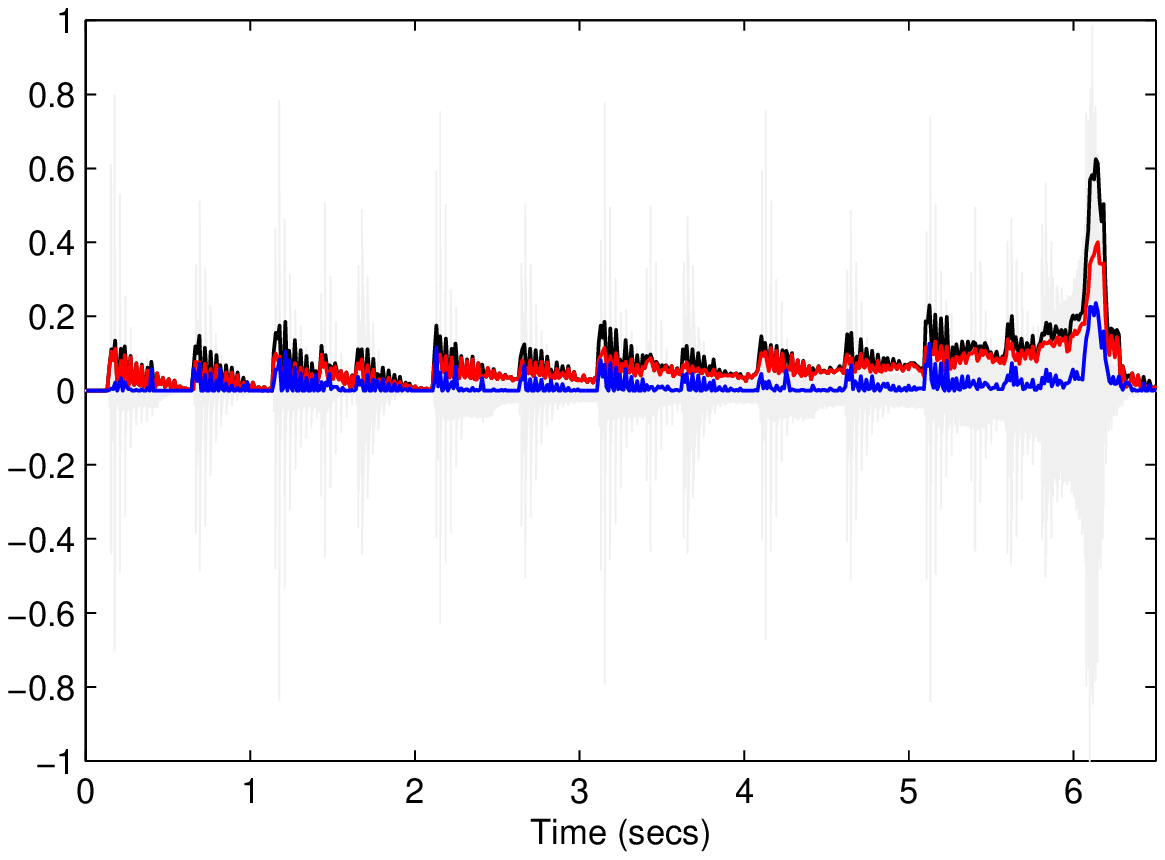}
\caption{\small{The dark line in the left 
graph joins the inverse local sparsity ratio 
values for the {\tt{Black\_Hole\_Billiards}} ring tone.
The lines in the right graph discriminate the 
inverse local sparsity ratio produced with atoms 
in the dictionary $\D_P$ (blue), in the dictionary $\D_T$ 
(red) and in the  whole dictionary $\D_M$ (black). 
\label{lsrb}}}
\end{center}
\end{figure}
 Since the  {\tt{Black\_Hole\_Billiards}} ring tone is 
 a more complex signal, due to the superposition of 
 the artificial sound, 
 the information given by the local 
sparsity ratio is richer than in the previous 
cases.
Notice for instance  that the dark 
line in the left graph of Fig.~\ref{lsrb} clearly indicates 
 the offsets in the percussive 
part of the clip which has been superimposed to the 
GS chirp.  Moreover this line, 
joining 512 points of inverse local sparsity ratio, 
also roughly follows the signal variation envelop.
 The graph on the right 
discriminates the local sparsity measure corresponding 
to atoms in the trigonometric component of the 
dictionary, and those in the  
dictionary $\D_P$. From bottom to top the first 
 line (blue) represents the inverse local sparsity values 
corresponding to atoms in $\D_P$ and the next line (red)
to atoms in $\D_T$. The top line (black) corresponds 
to atoms in the mixed dictionary $\D_M$ for facilitating 
the visual comparison. In this clip $20\%$ of atoms 
are from dictionary $\D_P$ and, as indicated by the blue line in the right graph of Fig.~\ref{lsrb}, a significant 
 contribution of those atoms takes place within the blocks 
where the rapid rise of the GS tone takes place 
(c.f. spectrogram in Fig.~\ref{bhb}).

\section{Conclusions} We have here advanced an 
effective technique for 
the numerical representation of Gravitational Sound 
clips produced by the 
 Laser Interferometer Gravitational-Wave Observatory (LIGO) 
 and the Massachusetts Institute of Technology (MIT). 
Our technique is inscribed   
  within the particular context of sparse representation 
and data reduction.
  We laid out a detailed procedure to this effect 
and were able to show that these types of signals can be 
approximated with high quality using  
{\em{significantly fewer elementary components}} 
 than those required 
within the standard orthogonal basis framework.

\subsection*{Acknowledgments} Thanks are due to 
LIGO, MIT and Milde Science Communications 
for making available the GS tones we have used 
in this paper. We are particularly grateful to 
Prof. S. A. Hughes and Prof. B. Schutz, 
for giving us information on 
 the generation of those signals.        

\end{document}